\documentclass[authoryear,preprint,review,12pt]{elsarticle}

\usepackage{amsmath, amssymb}
\usepackage{algorithm, algcompatible}
\usepackage{graphicx}
\usepackage[graphicx]{realboxes}
\usepackage{rotating}
\usepackage[english]{babel}

\usepackage{hyperref}
\hypersetup{
    colorlinks=true,
    linkcolor=blue,
    filecolor=magenta,      
    urlcolor=cyan,
    pdftitle={Overleaf Example},
    pdfpagemode=FullScreen,
    }
    
\usepackage{acro}   

\usepackage{lineno}

\usepackage{pgfplots}
\usepackage{pgf}
\usepackage{tikz}
\usetikzlibrary{matrix}

\usetikzlibrary{fit}					  
\usetikzlibrary{backgrounds}	          
\usetikzlibrary{shapes.arrows}
\usetikzlibrary{shapes.geometric}
\usetikzlibrary{shapes.multipart}
\usetikzlibrary{matrix, arrows, positioning}
\tikzstyle{rec}=[draw,rectangle, minimum height=2cm]
\tikzset{>=stealth', punkt/.style={
		fill=gray!40, 
		draw=black, very thick, text width=4.7em, minimum height=3em, text centered}}
\tikzset{>=stealth', Denoi/.style={rectangle, fill=blue!20, 
		draw=black, very thick, text width=6em, minimum height=3.5em, text centered}}
\tikzset{>=stealth', CG/.style={rectangle, 
		fill=green!20, draw=black, very thick, text width=9.5em, minimum height=3.5em, text centered}}

\tikzstyle{background} = [rectangle, fill=green!20, inner sep=0.1cm, rounded corners=4mm, 
minimum height=6em]
\tikzstyle{sum}   = [draw, fill=gray!40, circle, node distance=1cm]
\tikzstyle{dot}   = [circle, fill=black, inner sep=0pt, minimum size=5pt, node contents={}]
\tikzstyle{fig_n} = [node distance=30pt, inner sep=0cm]

\def\*#1{\mathbf{#1}}
\def\+#1{\mathcal{#1}}
\def\-#1{\mathbb{#1}}
\def\~#1{\mathrm{#1}}
\def\R{\mathbb{R}}

\def\Expect{\mathbb{E}}

\algnewcommand\INPUT{\item[\textbf{Input:}]}
\algnewcommand\OUTPUT{\item[\textbf{Output:}]}

\renewcommand{\hbar}{\bar{h}}

\journal{Computerized Medical Imaging and Graphics}

\begin{document}

\begin{frontmatter}

\title{Conditional Diffusion Models for CT Image Synthesis from CBCT: A Systematic Review}

\author{Alzahra Altalib\fnref{label1}, Chunhui Li\fnref{label2}, Alessandro Perelli\corref{cor1}\fnref{label3}}

\fntext[label1]{A. Altalib is with the Department of Applied Medical Sciences, Jordan University of Science and Technology, Irbid, 21410, Jordan, and the Department of Biomedical Engineering, University of Dundee, DD1 4HN, UK. A. Altalib is supported by the Jordan University of Science and Technology PhD scholarship.}
\fntext[label2]{C. Li is with the Department of Biomedical Engineering, University of Dundee, DD1 4HN, UK.}
\fntext[label3]{A. Perelli is with the School of Cardiovascular and Metabolic Health, College of Medicine, Veterinary and Life Sciences, University of Glasgow, G12 8TA, and with the Department of Biomedical Engineering, University of Dundee, DD1 4HN, UK. A. Perelli is supported by the Royal Academy of Engineering / Leverhulme Trust Research Fellowship LTRF-2324-20-160.}
\cortext[cor1]{Corresponding Author: Alessandro Perelli, email: \texttt{aperelli001@dundee.ac.uk}}

\begin{abstract}

\textit{Objective:} Cone-beam computed tomography (CBCT) provides a low-dose imaging alternative to conventional CT, but suffers from noise, scatter, and artifacts that degrade image quality. Synthetic CT (sCT) aims to translate CBCT to high-quality CT-like images for improved anatomical accuracy and dosimetric precision. Although deep learning approaches have shown promise, they often face limitations in generalizability and detail preservation. Conditional diffusion models (CDMs), with their iterative refinement process, offers a novel solution. This review systematically examines the use of CDMs for CBCT-to-sCT synthesis.

\textit{Methods:} A systematic search was conducted in Web of Science, Scopus, and Google Scholar for studies published between 2013 and 2024. Inclusion criteria targeted works employing conditional diffusion models specifically for sCT generation. Eleven relevant studies were identified and analyzed to address three questions: (1) What conditional diffusion methods are used? (2) How do they compare to conventional deep learning in accuracy? (3) What are their clinical implications?

\textit{Results:} CDMs incorporating anatomical priors and spatial-frequency features demonstrated improved structural preservation and noise robustness. Energy-guided and hybrid latent models enabled enhanced dosimetric accuracy and personalized image synthesis. Across studies, CDMs consistently outperformed traditional deep learning models in noise suppression and artefact reduction, especially in challenging cases like lung imaging and dual-energy CT.

\textit{Conclusion:} Conditional diffusion models show strong potential for generalized, accurate sCT generation from CBCT. However, clinical adoption remains limited. Future work should focus on scalability, real-time inference, and integration with multi-modal imaging to enhance clinical relevance.

\end{abstract}


\end{frontmatter}


\section{Introduction}\label{sec:Intro}

Cone-beam computed tomography (CBCT) and computed tomography (CT) are two of the widely used methods in clinical settings (\citet{hatcher2012ct}). The CBCT offers a low radiation method to provide real-time imaging and is a frequently used technique in image-guided radiotherapy. Despite having numerous applications, it suffers from increased noise, artifacts, and lower contrast for the soft tissues (\citet{schulze2011artefacts}). This may reduce its accuracy for the dose calculations and organ delineation. CT on the other side offer higher resolution. This results in a reliable Hounsfield Unit (HU) accuracy and the achievement of better soft tissue contrast. Thus CT is treated as a gold standard for treatment planning. Nevertheless, CT scans subject patients to higher radiation doses. Thus, patients cannot be subjected to the frequency of exposure to such scans. To cater to these limitations encountered by CT and CBCT, synthetic CT (sCT) generation methods are being used (\citet{fu2020cbct}). These methods take advantage of CBCT and CT and transform CBCT to high-quality sCT images which offer details at par with CT images. 
The sCT generation is traditionally carried out using deep learning methods. Two of the most widely used models in this context include generative adversarial networks (GANs) and variational autoencoders (VAEs) (\citet{chen2020synthetic}, \citet{liu2021gan}, \citet{zhang2022gan}). These methods have shown promising outcomes in the reduction of CBCT artifacts and improving the HU accuracy levels, however they face some challenges. Some of these challenges include mode collapse, limited structural fidelity, and higher dependency on the paired datasets. Additionally, the complexity inherently associated with medical images, especially that associated with capturing the fine anatomical details and noise, remains a gap to be addressed to attain high structural fidelity. Additionally, these challenges are required to be handled to provide improved generalizability across a diverse range of datasets. 
To overcome these limitations, diffusion models have emerged as alternative solutions for the sCT generation (\citet{peng2024conditional}). These methods offer several advantages such as handling the noise, preserving the structural details, and offering improved quantitative accuracy. The diffusion models are based on an iterative refinement strategy for the image denoising. This is followed by a reverse recovery process where images sCT images are generated to be closer to CT images. 
This review focuses on the exploration of conditional diffusion models and their applications in the sCT generation. The model evaluation of these methods in light of performance and clinical applications has been carried out. A synthesis from the recent studies is aimed to be carried out by analyzing the advancements made by the diffusion models and their advantages over the traditional models. Ideally, the review analyzes the strength of diffusion models in addressing existing research gaps in the sCT generation for improved patient outcomes.

\section{Overview of Diffusion Model Families for CT Image Synthesis}

Diffusion-based synthetic CT generation begins with learning a mapping from noisy or artifact-prone CBCT images to clean, diagnostic-quality sCT representations. The target is to recover a high-resolution, anatomically aligned CT image that retains structural fidelity to the CBCT input. Although earlier methods explored GAN-based mappings, their limitations in uncertainty modeling and structural stability have made diffusion-based techniques more attractive.

Diffusion models for CBCT-to-sCT are typically formulated as conditional generative models, where the CBCT serves as conditioning input. This enables patient-specific image generation with consistent anatomical correspondence. High spatial and contrast fidelity is required to ensure accurate downstream use in radiotherapy dose calculation or treatment planning. Thus, generative models must respect both global and local anatomical context present in CBCT, although learning to correct for modality-specific limitations such as noise and scatter.

\subsection{CT Output Specificity in CBCT-conditioned Models}

The output specificity of sCT generation models is of paramount clinical importance. For sCT to be usable in dose computation or image-guided radiotherapy, the generated CT must not only appear realistic but must also be anatomically accurate with respect to the CBCT. Diffusion models provide several advantages here:

\begin{itemize}
    \item Structural Preservation: The denoising process is inherently robust to noise perturbations, and when strongly conditioned on CBCT, the model tends to preserve macro and micro anatomical features, including subtle tissue boundaries.

    \item Uncertainty Modeling: Unlike deterministic models, diffusion models model a posterior distribution over CT outputs conditioned on CBCT. This allows clinicians to assess confidence levels and potentially detect ambiguous or low-quality regions.

    \item Multi-modal Consistency: When paired with frequency-domain or multi-scale conditioning (e.g., FGDA), the model can enforce consistency between CT texture and CBCT geometry, thus increasing diagnostic confidence.
\end{itemize}

\noindent Output specificity can be quantitatively assessed using metrics such as structural similarity (SSIM), peak signal-to-noise ratio (PSNR), and clinical dose deviation. However, visual and expert-based evaluations remain crucial due to the clinical significance of subtle anatomical discrepancies.

In summary, strong CBCT conditioning combined with diffusion-based generative modeling yields high-specificity synthetic CT images suitable for integration into clinical workflows.

\section{Analysis of Diffusion Model Variants}

In the following sections, we analyze four prominent diffusion model variants, denoising diffusion probabilistic models (DDPMs), denoising diffusion implicit models (DDIMs), latent diffusion models (LDMs), and frequency-guided diffusion models (FGDMs), within the context of CBCT-conditioned synthetic CT generation. Our exposition emphasizes both theoretical underpinnings and clinical implications.


\subsection{Denoising Diffusion Probabilistic Models (DDPMs)}
DDPMs constitute the foundational class of diffusion models. They simulate a forward process in which Gaussian noise is gradually added to an image over multiple timesteps and a reverse process that learns to denoise this corrupted data iteratively, ultimately recovering the clean image distribution from noise (\citet{ho2020denoising}). The DDPM approach is formally grounded in variational inference and models the data likelihood through a series of conditional Gaussians. 

Let $\*x_0 \in \R^{H \times W}$ denote the ground truth CT image, and let $\*y$ be the conditioning CBCT input image. The goal is to learn a conditional distribution $p_\theta(\*x_0 \mid \*y)$ that generates synthetic CT images $\*x_0$ conditioned on $\*y$.

A DDPM defines a noising process $q$ that gradually adds Gaussian noise to data over $T$ steps. The process is defined as:

\begin{eqnarray}
    q(\*x_{1:T} \mid \mathbf{x}_0) & = & \prod_{t=1}^T q(\*x_t \mid \*x_{t-1}), \\
    q(\*x_t \mid \*x_{t-1}) & := & \+N(\*x_t; \sqrt{1 - \beta_t} \*x_{t-1}, \beta_t \*I) \nonumber 
\end{eqnarray}
where $\{\beta_t\}_{t=1}^T$ is a variance schedule, typically linear or cosine. The marginal distribution of $\*x_t$ given $\*x_0$ can be derived analytically:

\begin{equation}
    q(\*x_t \mid \*x_0) = \mathcal{N}(\*x_t; \sqrt{\bar{\alpha}_t} \*x_0, (1 - \bar{\alpha}_t) \*I),
\end{equation}
where $\alpha_t := 1 - \beta_t$ and $\bar{\alpha}_t := \prod_{s=1}^t \alpha_s$. This allows one to sample $\*x_t \sim q(\*x_t \mid \*x_0)$ directly.

\noindent The reverse (generative) process is another Markov chain parameterized by a neural network:

\begin{eqnarray}
    p_\theta(\hat{\*x}_{0:T} \mid \*y) & = & p(\*x_T) \prod_{t=1}^T p_\theta(\hat{\*x}_{t-1} \mid \hat{\*x}_t, \*y), \\
    p_\theta(\hat{\*x}_{t-1} \mid \hat{\*x}_t, \*y) & := & \mathcal{N}(\hat{\*x}_{t-1}; \boldsymbol{\mu}_\theta(\hat{\*x}_t, t, \*y), \Sigma_\theta(\hat{\*x}_t, t, \*y)) \nonumber
\end{eqnarray}

In most implementations, the variance $\Sigma_\theta$ is either fixed or learned separately. A common simplification uses:

\begin{equation}
\boldsymbol{\mu}_\theta(\hat{\*x}_t, t, \*y) = \frac{1}{\sqrt{\alpha_t}} \left( \hat{\*x}_t - \frac{\beta_t}{\sqrt{1 - \bar{\alpha}_t}} \boldsymbol{\epsilon}_\theta(\hat{\*x}_t, t, \*y) \right),
\end{equation}
where $\boldsymbol{\epsilon}_\theta$ predicts the noise added at step $t$.

\noindent The training loss is based on the variational lower bound (VLB) on the conditional negative log-likelihood which is:

\begin{equation}
\log p_\theta(\*x_0 \mid \*y) \geq \Expect_{q} \left[ \log \frac{p_\theta(\*x_{0:T} \mid \*y)}{q(\*x_{1:T} \mid \*x_0)} \right] =: -\+L_\text{VLB}
\end{equation}

This decomposes into per-timestep KL divergences:

\begin{equation}
\mathcal{L}_\text{VLB} = \Expect_q \left[ \sum_{t=1}^T D_\text{KL}(q(\*x_{t-1} \mid \*x_t, \*x_0) \parallel p_\theta(\hat{\*x}_{t-1} \mid \hat{\*x}_t, \mathbf{y})) - \log p_\theta(\hat{\*x}_0 \mid \hat{\*x}_1, \*y) \right]
\end{equation}

\noindent A practical surrogate loss simplifies the training objective to:

\begin{equation}
\mathcal{L}_\text{simple} = \Expect_{\*x_0, \boldsymbol{\epsilon}, t} \left[ \left\| \boldsymbol{\epsilon} - \boldsymbol{\epsilon}_\theta(\sqrt{\bar{\alpha}_t} \*x_0 + \sqrt{1 - \bar{\alpha}_t} \boldsymbol{\epsilon}, t, \*y) \right\|^2 \right],
\end{equation}
where $\boldsymbol{\epsilon} \sim \mathcal{N}(0, \mathbf{I})$. This is essentially denoising score matching.

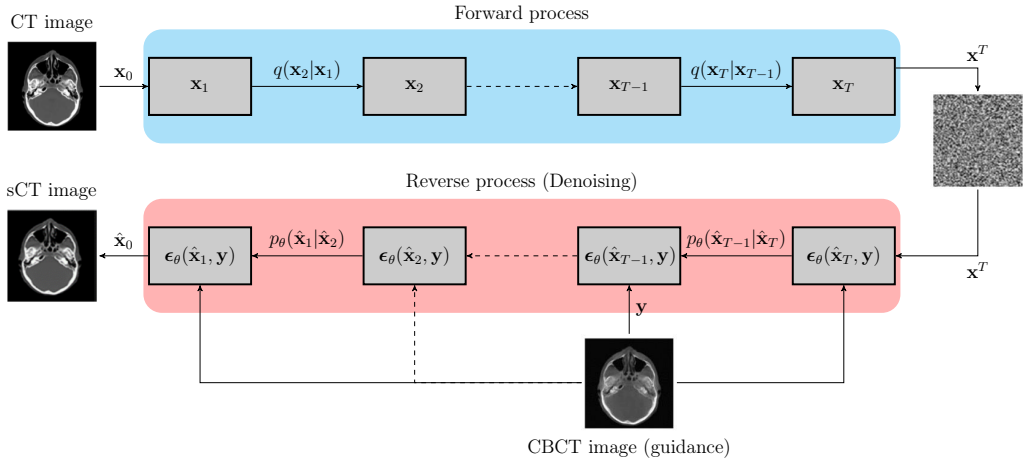
\begin{figure*}[!h]
	\resizebox{\textwidth}{!}{
        \centering
		\begin{tikzpicture}[node distance=2.4cm,scale=1]
			\node [punkt] (x1) {};
			\node[fig_n, left= of x1, inner sep=0pt, label=above:CT image] (IN) 
			{\includegraphics[width=.15\textwidth]{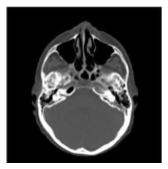}};	
            \node [punkt] (x1) {$\*x_1$};
			\path[->] (IN.east) edge node [above] {$\*x_0$} (x1.west);
			\node [punkt, right= of x1] (x2) {$\*x_2$};
			\path[->] (x1.east) edge node [above] {$q(\*x_2 | \*x_1)$} (x2.west);
			
			\node [punkt, right= of x2] (xt_1) {$\*x_{T-1}$};
			\path[->, dashed] (x2.east) edge node [above] {} (xt_1.west);
			\node [punkt, right= of xt_1] (xt) {$\*x_T$};
			\path[->] (xt_1.east) edge node [above] {$q(\*x_T | \*x_{T-1})$} (xt.west);
			
			\node[fig_n, below right= of xt.north east, inner sep=0pt] (noise) 
			{\includegraphics[width=.15\textwidth]{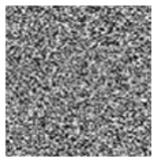}};	
			\draw[->] (xt.east)+(0,0.4) -| (noise.north) node [midway, above] {$\*x^T$};
			
			\node [punkt, below= of xt] (xtr) {$\boldsymbol{\epsilon}_\theta(\hat{\*x}_T, \*y)$};
			\draw[->] (noise.south) |- (xtr.east)  node [midway, below] {$\*x^T$};
			\node [punkt, left= of xtr] (xt_1r) {$\boldsymbol{\epsilon}_\theta(\hat{\*x}_{T-1}, \*y)$};
			\path[->] (xtr.west) edge node [above] {$p_{\theta}(\hat{\*x}_{T-1} | \hat{\*x}_T)$} (xt_1r.east);
			\node [punkt, left= of xt_1r] (x2r) {$\boldsymbol{\epsilon}_\theta(\hat{\*x}_2, \*y)$};
			\node [punkt, left= of x2r] (x1r) {$\boldsymbol{\epsilon}_\theta(\hat{\*x}_1, \*y)$};
			\path[->, dashed] (xt_1r.west) edge node [above] {} (x2r.east);
			\path[->] (x2r.west) edge node [above] {$p_{\theta}(\hat{\*x}_1 | \hat{\*x}_2)$} (x1r.east);
			\node[fig_n, left= of x1r, inner sep=0pt, label=above:sCT image] (sCT) {\includegraphics[width=.15\textwidth]{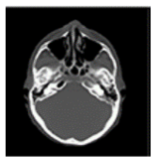}};	
			\path[->] (x1r.west) edge node [above] {$\hat{\*x}_0$} (sCT.east);
			
			\node[fig_n, below= of xt_1r, inner sep=0pt, label=below:CBCT image (guidance)] (CBCT) {\includegraphics[width=.15\textwidth]{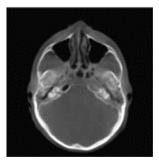}};	
			\path[->] (CBCT.north) edge node [right] {$\*y$} (xt_1r.south);
			\draw [->] (CBCT.east) -| (xtr.south) [right] node{};
			\draw [->, dashed] (CBCT.west) -| (x2r.south) [right] node{};
			\draw [->] (CBCT.west) -| (x1r.south) [right] node{};
			
			\begin{pgfonlayer}{background}[node distance=1cm]
				\node [background, fit=(x1) (x2) (xt_1) (xt), fill=cyan!30, label=above:Forward process] (p_decomp) {};
			\end{pgfonlayer}
			\begin{pgfonlayer}{background}[node distance=1cm]
				\node [background,fit=(xtr) (xt_1r) (x2r) (x1r), fill=red!30, label=above:Reverse process (Denoising)] (m_denoi) {};
			\end{pgfonlayer} 
	\end{tikzpicture}}
	\caption{Workflow of Conditional DDPM for CBCT to CT image synthesis.}\label{fig:DDPM_workflow}
\end{figure*}

This approach has been depicted in Fig. \ref{fig:DDPM_workflow}  and allows them to reconstruct high-quality sCT images with improved artefact reduction and HU accuracy. 


\noindent At inference, to generate an sCT image it involves the reverse generative process as in Algorithm \ref{table:DDPM}. 

\begin{algorithm}[!h]
	\caption{DDPM algorithm}
	\label{table:DDPM}
	\begin{algorithmic}[1]
		\REQUIRE Sample $\*x_T \sim \+N(0, \*I)$.
		\FOR{$t = T, \ldots, 1$}
		\STATE $\hat{\*x}_{t-1} \sim \+N\left(\boldsymbol{\mu}_\theta(\hat{\*x}_t, t, \*y), \Sigma_t\right)$
		\OUTPUT $\*x_0$
		\ENDFOR
	\end{algorithmic}
\end{algorithm} 
		

This process is stochastic and allows for sampling multiple plausible sCTs per input $\*y$.

\subsection{Conditioning Mechanisms in CBCT-to-sCT Diffusion Models}

As mentioned above, conditioning the diffusion process on CBCT input is central to the success of sCT generation. The goal is to guide the generative process so that the output is not only plausible as a CT image, but also accurately reflects the anatomical structure present in the CBCT. Several strategies have been explored in the literature:

\begin{itemize}
    \item Input Concatenation: The CBCT volume is concatenated with the noisy latent or image sample at each denoising step. Often it is implemented using a conditional UNet where  $\boldsymbol{\epsilon}_\theta(\hat{\*x}_t, t, \*y)$ takes $\*y$ as input, typically via concatenation. This direct approach has the benefit of simplicity and early integration of structural information. 

    \item Feature Modulation (FiLM): The CBCT input $\*y$ is encoded via a convolutional network, and its features are used to modulate the internal layers of the denoising network. FiLM layers apply affine transformations conditioned on CBCT features, allowing spatial and channel-specific influence. This effectively models $p_\theta(\*x_0 \mid \*y)$ without needing explicit paired supervision.

    \item Cross-Attention: Particularly powerful in vision transformers and UNet-based diffusion models, cross-attention enables the model to selectively integrate information from CBCT across spatial scales. The attention maps provide interpretability and allow flexible registration-free alignment.

    \item Classifier-Free Guidance (CFG): This strategy trains the diffusion model with and without conditioning. During inference, guidance strength is controlled by interpolating between the two outputs. 
    This enhances fidelity by adjusting conditional vs unconditional noise estimates:

    \begin{equation}
        \boldsymbol{\epsilon}_{\text{guided}} = (1 + w) \boldsymbol{\epsilon}_\theta(\hat{\*x}_t, t, \*y) - w \boldsymbol{\epsilon}_\theta(\hat{\*x}_t, t),
    \end{equation}
    where $w > 0$ is a guidance scale.
    
    This allows tuning the influence of CBCT during generation, which is especially useful when dealing with varying levels of CBCT quality.
\end{itemize}

\noindent Each of these mechanisms seeks to achieve anatomical fidelity although allowing flexibility in how much and where CBCT information is used. For sCT applications, high-resolution alignment, especially in bone and soft-tissue boundaries, is critical. Thus, architectural choices that facilitate multiscale fusion of CBCT features are preferred.

The advantages of the DDPM is that it captures uncertainty since output diversity reflects aleatoric uncertainty in sCT generation and it is a non-adversarial training approach which is more stable compared to GANs. Hovewer, the sampling might be slow since it requires  $T \sim 1000$ and large compute resources for training and inference. Furthermore, it is of outmost importance to align CT and CBCT conditioning feature. Overall although DDPMs yield high-fidelity outputs, they are computationally intensive due to the large number of sequential denoising steps required.

\subsection{Denoising Diffusion Implicit Models (DDIMs)}
DDIMs redefine the generative process as a non-Markovian, deterministic mapping that preserves the same training objective as DDPM but modifies the sampling process, allowing fewer sampling steps without significant loss in quality (\citet{song2020denoising}). By leveraging a reparameterized trajectory through the diffusion space, DDIMs enable faster inference.  

Given a noisy sample $\*x_t$, we deterministically obtain $\mathbf{x}_{t-1}$ via:

\begin{equation}
    \hat{\*x}_{t-1} = \sqrt{\bar{\alpha}_{t-1}} \*x_0 + \sqrt{1 - \bar{\alpha}_{t-1} - \eta^2 (1 - \bar{\alpha}_t) / (1 - \bar{\alpha}_{t-1})} \boldsymbol{\epsilon}_\theta(\hat{\*x}_t, t, \*y) + \eta \boldsymbol{\epsilon},
\end{equation}
where $\eta \in [0, 1]$ controls the stochasticity ($\eta = 0$ yields a fully deterministic path). In the most common setting:

\begin{equation}
    \hat{\*x}_{t-1} = \sqrt{\bar{\alpha}_{t-1}} \left( \frac{\hat{\*x}_t - \sqrt{1 - \bar{\alpha}_t} \boldsymbol{\epsilon}_\theta(\hat{\*x}_t, t, \*y)}{\sqrt{\bar{\alpha}_t}} \right) + \sqrt{1 - \bar{\alpha}_{t-1}} \boldsymbol{\epsilon}_\theta(\hat{\*x}_t, t, \*y).
\end{equation}

This eliminates the need to sample Gaussian noise $\boldsymbol{\epsilon}$ during generation, drastically accelerating inference (e.g., from 1000 to 50 steps) without retraining.

Although not always emphasized in medical imaging tasks, DDIMs provide a practical trade-off between speed and image quality, making them suitable for real-time or interactive applications. 

\subsection{Latent Diffusion Models (LDMs)}
Latent diffusion models address the computational burden of pixel-space generation by operating in a compressed latent space learned via autoencoders or variational encoders (VAE) (\citet{rombach2022high}). By learning the diffusion process in this low-dimensional domain, LDMs dramatically lower the memory and runtime cost, enabling the use of higher-resolution medical data. Once denoising is completed in the latent domain, a decoder reconstructs the final image. These models are the class of generative models that can work and operate on the low dimensional latent space instead of the direct image space. LDMs make use of VAE-based encoder-decoder settings for learning the compressed latent representation of the CT images. 
A CT image $\*x_0$ is first encoded into a latent vector $\*z_0$ using a VAE encoder $\+E(\cdot)$: 
\begin{equation}
    \*z_0 = \+E(\*x_0)
\end{equation}
and the forward diffusion process operates in Latent Space where the noise is progressively added to the latent vector $\*z_0$ to obtain a noisy latent vector $\*z_t$ at time step $t$:

\begin{equation}
    \*z_t=\sqrt{\bar{\alpha}_t} \*z_0 + \sqrt{1 - \bar{\alpha}_t}\boldsymbol{\epsilon}, \quad \boldsymbol{\epsilon}\sim\+N(0,\*I)
\end{equation}

In the reverse process the CT image generation is conditioned on the CBCT image $\*y$ which is passed through a Condition Encoder to generate a context vector used to guide the reverse diffusion:
\begin{equation}
    \*z_{\*y} = \+E_{\*y} (\*y)
\end{equation}
This conditional embedding captures anatomical priors for guiding the reverse denoising process. 

In the reverse process the noisy latent vector $\*z_t$ at timestep $t$, is used predicts the noise component $\boldsymbol{\epsilon}_t$ using the and the condition embedding $\*z_{\*y}$:

\begin{equation}
    \boldsymbol{\epsilon}_t=\boldsymbol{\epsilon}_{\theta} (\hat{\*z}_t, t, \*z_{\*y})
\end{equation}
The DDIM-style update is then applied:
\begin{equation}
    \hat{\*z}_{t-1}=\sqrt{\bar{\alpha}_{t-1}} \left( \frac{\hat{\*z}_t - \sqrt{1 - \bar{\alpha}_t}\boldsymbol{\epsilon}_t}{\bar{\alpha}_t} \right) + \sqrt{1 - \bar{\alpha}_{t-1}}\boldsymbol{\epsilon}_t
\end{equation}

After denoising the latent back to $\*z_0$, the final sCT image is reconstructed using the VAE decoder $\+D(\cdot)$:
\begin{equation}
    \hat{\*x}_0 = \+D(\hat{\*z}_0)
\end{equation}

\begin{figure*}[!h]
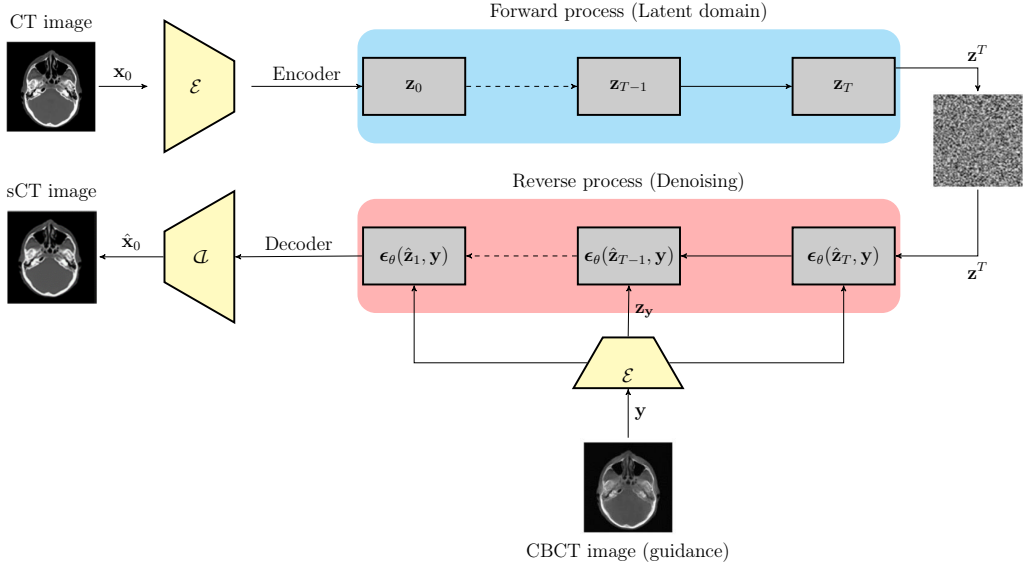

	\resizebox{\textwidth}{!}{
        \centering
		\begin{tikzpicture}[node distance=2.4cm,scale=1]
			\node [punkt, fill=yellow!30, trapezium, rotate=270, align=center, text width=2em, minimum height=1.2em, text height=3em] (x1) (0, -1.5) {\rotatebox{90}{$\quad\+E$}};
			\node[fig_n, left= of x1, inner sep=0pt, label=above:CT image] (IN) 
			{\includegraphics[width=.15\textwidth]{images/CT.png}};	
            
			\path[->] (IN.east) edge node [above] {$\*x_0$} (x1.west);
			\node [punkt, right= of x1] (x2) {$\*z_0$};
			\path[->] (x1.east) edge node [above] {Encoder} (x2.west);
			
			\node [punkt, right= of x2] (xt_1) {$\*z_{T-1}$};
			\path[->, dashed] (x2.east) edge node [above] {} (xt_1.west);
			\node [punkt, right= of xt_1] (xt) {$\*z_T$};
			\path[->] (xt_1.east) edge node [above] {} (xt.west);
			
			\node[fig_n, below right= of xt.north east, inner sep=0pt] (noise) 
			{\includegraphics[width=.15\textwidth]{images/noise.jpeg}};	
			\draw[->] (xt.east)+(0,0.4) -| (noise.north) node [midway, above] {$\*z^T$};
			
			\node [punkt, below= of xt] (xtr) {$\boldsymbol{\epsilon}_\theta(\hat{\*z}_T, \*y)$};
			\draw[->] (noise.south) |- (xtr.east)  node [midway, below] {$\*z^T$};
			\node [punkt, left= of xtr] (xt_1r) {$\boldsymbol{\epsilon}_\theta(\hat{\*z}_{T-1}, \*y)$};
			\path[->] (xtr.west) edge node [above] {} (xt_1r.east);
			\node [punkt, left= of xt_1r] (x2r) {$\boldsymbol{\epsilon}_\theta(\hat{\*z}_1, \*y)$};
			\node [punkt, fill=yellow!30, trapezium, rotate=90, text width=2em, minimum height=1.2em, text height=3em] (x1r) at (0, -3.7) {\rotatebox{90}{$\quad\+D$}};
			\path[->, dashed] (xt_1r.west) edge node [above] {} (x2r.east);
			\path[->] (x2r.west) edge node [above] {Decoder} (x1r.south);
			\node[fig_n, inner sep=0pt, label=above:sCT image] (sCT) at (-3.2, -3.7) {\includegraphics[width=.15\textwidth]{images/sCT.png}};	
			\path[->] (x1r.north) edge node [above] {$\hat{\*x}_0$} (sCT.east);

            \node [punkt, fill=yellow!30, trapezium, align=center, text width=2em, minimum height=1em, text height=2em] (Ey) at (9.3, -6) {$\+E$};
            
			\node[fig_n, below= of Ey, inner sep=0pt, label=below:CBCT image (guidance)] (CBCT) {\includegraphics[width=.15\textwidth]{images/CBCT.jpeg}};	
			\path[->] (Ey.north) edge node [right] {$\*z_{\*y}$} (xt_1r.south);
			\draw [->] (Ey.east) -| (xtr.south) [right] node{};
			\draw [->] (Ey.west) -| (x2r.south) [right] node{};
			\path[->] (CBCT.north) edge node [right] {$\*y$} (Ey.south);
			
			\begin{pgfonlayer}{background}[node distance=1cm]
				\node [background, fit=(x2) (xt_1) (xt), fill=cyan!30, label=above:Forward process (Latent domain)] (p_decomp) {};
			\end{pgfonlayer}
			\begin{pgfonlayer}{background}[node distance=1cm]
				\node [background,fit=(xtr) (xt_1r) (x2r), fill=red!30, label=above:Reverse process (Denoising)] (m_denoi) {};
			\end{pgfonlayer} 
	\end{tikzpicture}}
	\caption{Workflow of LDM for CT image synthesis using CBCT guidance.}\label{fig:LDM_workflow}
\end{figure*}

\noindent The diagram in Fig. \ref{fig:LDM_workflow} show the training workflow of the forward and reverse diffusion processes in the latent domain with conditional CBCT guidance.  

\subsection{Frequency-Guided Diffusion Models (FGDMs)}
FGDMs introduce frequency-domain priors to diffusion-based generation, with the aim of improving the recovery of high-frequency details often lost in noisy imaging modalities (\citet{Li_2024}). These models typically embed frequency-aware losses or frequency-decomposed guidance into the diffusion pipeline, enabling sharper reconstruction of anatomical boundaries and fine textures. For CBCT applications, FGDMs are particularly effective in restoring bone edges and soft tissue transitions, which are otherwise blurred due to scatter and noise.

FGDA aids in improving the diffusion process through an incorporation of spatial and frequency-domain information. The idea is to denoise the network and frequency-domain features which are extracted from the CBCT images. This process leads to an improved anatomical details and fidelity in the sCT reconstruction. 
As in standard diffusion models, the forward process perturbs a clean CT image $\*x_0$ into a noisy version $\*x_T$ by sequentially adding Gaussian noise:
\begin{equation}
    \*x_t=\sqrt{\bar{\alpha}_t} \*x_0 + \sqrt{1 - \bar{\alpha}_t}\boldsymbol{\epsilon}, \quad \boldsymbol{\epsilon}\sim \+N(0,\*I),\quad t=1, \ldots, T
\end{equation}

To guide the reverse process, a frequency-domain representation $f$ of the image is extracted by applying a Frequency Transform (e.g., DCT or FFT $\+F$) to the CBCT image $\*Y$ to obtain frequency coefficients. A frequency Encoder $\+E_f(\cdot)$ and condition Encoder $\+E_c(\cdot)$ are used to derive rich feature representations. To guide the reverse process a combined conditioning vector is obtained by concatenating spatial and frequency-based features:
\begin{equation}
    \*z_{\*f} = \+E_f(\+F(\*y)) \oplus \+E_c(\*y) 
\end{equation}
where $\+F(\cdot)$ denotes the frequency transform and $\oplus$ indicated the vector concatenation. 

The reverse process is carried out using a noise predictor $\boldsymbol{\epsilon}_{\theta}$ guided by both spatial and frequency information:

\begin{equation}
    \boldsymbol{\epsilon}_t=\boldsymbol{\epsilon}_{\theta} (\hat{\*x}_t, t, \*z_{\*f})
\end{equation}
This is used in the deterministic update rule (as in DDIM): 
\begin{equation}
    \hat{\*x}_{t-1}=\sqrt{\bar{\alpha}_{t-1}} \left( \frac{\hat{\*x}_t - \sqrt{1 - \bar{\alpha}_t}\boldsymbol{\epsilon}_t}{\bar{\alpha}_t} \right) + \sqrt{1 - \bar{\alpha}_{t-1}}\boldsymbol{\epsilon}_t
\end{equation}

\begin{figure}[!h]
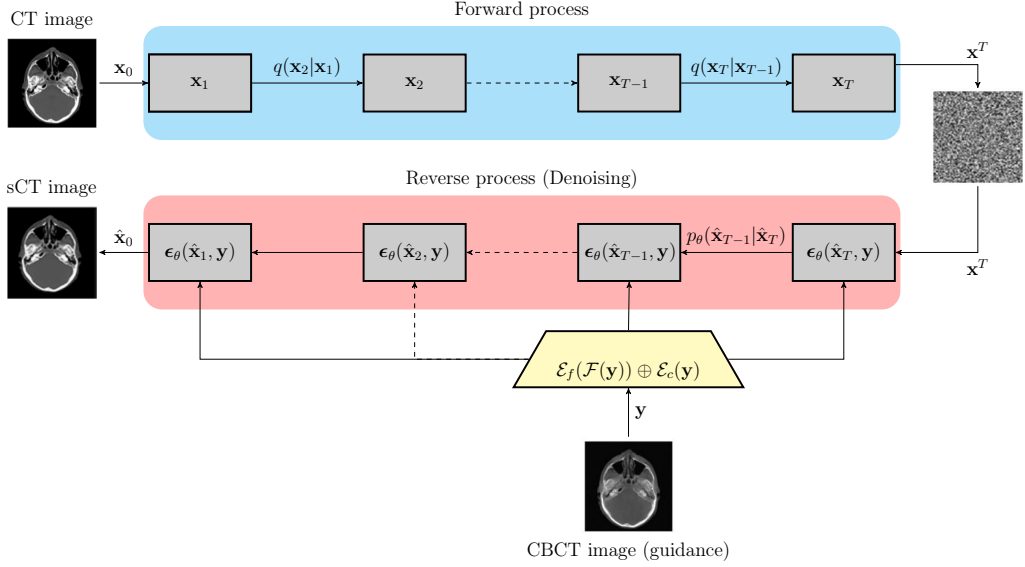

	\resizebox{\textwidth}{!}{
        \centering
		\begin{tikzpicture}[node distance=2.4cm,scale=1]
			\node [punkt] (x1) {};
			\node[fig_n, left= of x1, inner sep=0pt, label=above:CT image] (IN) 
			{\includegraphics[width=.15\textwidth]{images/CT.png}};	
            \node [punkt] (x1) {$\*x_1$};
			\path[->] (IN.east) edge node [above] {$\*x_0$} (x1.west);
			\node [punkt, right= of x1] (x2) {$\*x_2$};
			\path[->] (x1.east) edge node [above] {$q(\*x_2 | \*x_1)$} (x2.west);
			
			\node [punkt, right= of x2] (xt_1) {$\*x_{T-1}$};
			\path[->, dashed] (x2.east) edge node [above] {} (xt_1.west);
			\node [punkt, right= of xt_1] (xt) {$\*x_T$};
			\path[->] (xt_1.east) edge node [above] {$q(\*x_T | \*x_{T-1})$} (xt.west);
			
			\node[fig_n, below right= of xt.north east, inner sep=0pt] (noise) 
			{\includegraphics[width=.15\textwidth]{images/noise.jpeg}};	
			\draw[->] (xt.east)+(0,0.4) -| (noise.north) node [midway, above] {$\*x^T$};
			
			\node [punkt, below= of xt] (xtr) {$\boldsymbol{\epsilon}_\theta(\hat{\*x}_T, \*y)$};
			\draw[->] (noise.south) |- (xtr.east)  node [midway, below] {$\*x^T$};
			\node [punkt, left= of xtr] (xt_1r) {$\boldsymbol{\epsilon}_\theta(\hat{\*x}_{T-1}, \*y)$};
			\path[->] (xtr.west) edge node [above] {$p_{\theta}(\hat{\*x}_{T-1} | \hat{\*x}_T)$} (xt_1r.east);
			\node [punkt, left= of xt_1r] (x2r) {$\boldsymbol{\epsilon}_\theta(\hat{\*x}_2, \*y)$};
			\node [punkt, left= of x2r] (x1r) {$\boldsymbol{\epsilon}_\theta(\hat{\*x}_1, \*y)$};
			\path[->, dashed] (xt_1r.west) edge node [above] {} (x2r.east);
			\path[->] (x2r.west) edge node [above] {} (x1r.east);
			\node[fig_n, left= of x1r, inner sep=0pt, label=above:sCT image] (sCT) {\includegraphics[width=.15\textwidth]{images/sCT.png}};	
			\path[->] (x1r.west) edge node [above] {$\hat{\*x}_0$} (sCT.east);

            \node [punkt, fill=yellow!30, trapezium, align=center, text width=8em, minimum height=1em, text height=2em] (Ey) at (9.3, -6) {$\+E_f(\+F(\*y)) \oplus \+E_c(\*y)$ };
            
		  \node[fig_n, below= of Ey, inner sep=0pt, label=below:CBCT image (guidance)] (CBCT) {\includegraphics[width=.15\textwidth]{images/CBCT.jpeg}};	
			\path[->] (CBCT.north) edge node [right] {$\*y$} (Ey.south);
            \path[->] (Ey.north) edge node [right] {} (xt_1r.south);
			\draw [->] (Ey.east) -| (xtr.south) [right] node{};
			\draw [->, dashed] (Ey.west) -| (x2r.south) [right] node{};
			\draw [->] (Ey.west) -| (x1r.south) [right] node{};
			
			\begin{pgfonlayer}{background}[node distance=1cm]
				\node [background, fit=(x1) (x2) (xt_1) (xt), fill=cyan!30, label=above:Forward process] (p_decomp) {};
			\end{pgfonlayer}
			\begin{pgfonlayer}{background}[node distance=1cm]
				\node [background,fit=(xtr) (xt_1r) (x2r) (x1r), fill=red!30, label=above:Reverse process (Denoising)] (m_denoi) {};
			\end{pgfonlayer} 
	\end{tikzpicture}}
	\caption{Workflow of the Frequency Guided Diffusion model for CT image Synthesis.}\label{fig:FGDM_workflow}
\end{figure}

\noindent The diagram in Fig. \ref{fig:FGDM_workflow} show the training workflow of the diffusion process with conditional CBCT frequency guidance.

\section{Aim of the Study}\label{sec:aim}
The primary aim of this systematic review is to critically evaluate the application of conditional diffusion models for generating synthetic CT (sCT) images from cone-beam CT (CBCT) data. Specifically, this study seeks to address the limitations of CBCT, such as image noise, scatter, and artifacts, by exploring how diffusion models can improve the quality and clinical utility of sCT. The review aims to systematically identify the methodologies employed in conditional diffusion approaches, compare their performance with traditional deep learning techniques in terms of accuracy and robustness, and examine their clinical relevance. Ultimately, the study aims to provide insights into the potential of conditional diffusion models to enhance dosimetric precision and anatomical fidelity in radiotherapy, although  highlighting gaps and directions for future research.

\subsection{Research Questions}\label{subsec:r_quest}

\begin{enumerate}
    \item Which of the methods in conditional diffusion models for sCT are employed? 
    \item How are diffusion models compared to traditional deep learning models in terms of accuracy? 
    \item What are the clinical implications of using diffusion models for sCT generation?
\end{enumerate}

\section{Methodology}\label{sec:method}
This section begins by outlining the search strategy employed to identify relevant studies for the review. Following this, a comprehensive description of the systematic and methodical steps undertaken to conduct the review is provided, ensuring transparency and replicability of the research process.

\subsection{Search Strategy:}
This systematic search followed the PRISMA statement (\citet{page2021prisma}) and used the PICO model Table \ref{tab:pico} to find relevant literature. PubMed, Web of Science (WOS), Scopus, IEEE Xplore, and Google Scholar databases were searched from 2013- 2024, following the defined criteria of the study, to ensure the inclusion of all pertinent studies. The search strategy utilized a combination of phrases and keywords relevant to the research question to guarantee comprehensive coverage. These included ' Diffusion Model ', ' Conditional Diffusion ', 'cone beam computed tomography', 'dose calculation,' and synonyms such as 'CBCT.' Boolean operators (AND, OR)" were utilized to search for different database appendices to efficiently filter and merge search keywords.

\begin{table}[!h]
\centering
\renewcommand{\arraystretch}{1.5}
\setlength{\tabcolsep}{10pt}
\caption{The PICO framework for systematic reviews.}
\begin{tabular}{|l|p{9.7cm}|}
\hline
\textbf{Population} & All patients underwent definitive oncology planning. \\
\hline
\textbf{Intervention} & Diffusion Model OR Conditional Diffusion OR Denoising Diffusion OR Score-Based Generative Model, CBCT OR Cone Beam Computed Tomography OR Cone-Beam CT, Imaging Reconstruction OR IR, Unsupervised Deep Learning OR UDL, Dose Estimations OR DE, Medical Imaging OR MI, Artifact Reduction OR AR, Radiotherapy. \\
\hline
\textbf{Comparison} & CT OR Computed Tomography. \\
\hline
\textbf{Outcome} & Sensitivity, Specificity, Accuracy. \\
\hline
\end{tabular}
\label{tab:pico}
\end{table}

\subsection{Inclusion Criteria}

\begin{enumerate}
\item Articles explicitly addressing diffusion models for synthetic CT generation.
\item Research employing conditional approaches in diffusion models, such as guidance by specific features or anatomical priors.
\item Peer-reviewed journal articles, conference proceedings, and systematic reviews related to diffusion-based synthetic imaging.
\item Publications from the last 11 years (2013–2024).
\item Articles published in English.
\end{enumerate}

\subsection{Exclusion Criteria}

\begin{enumerate}
\item Studies do not involve diffusion models as a primary method for synthetic CT generation.
\item Papers on CBCT enhancement, noise reduction, or artefact correction unrelated to synthetic CT generation.
\item Non-peer-reviewed articles such as blogs, editorials, or opinion pieces.
\item Duplicates across databases.
\item Articles lacking sufficient methodological details or evaluation metrics.
\item Articles published in non-English languages or without reliable translations.
\item Publications before 2013 unless they are foundational studies in diffusion models.
\end{enumerate}

\subsection{Study Selection:}
Following the systematic search, articles are screened based on their titles and abstracts to identify potentially relevant studies· After undergoing an initial screening process, articles proceed to a full-text review, where their suitability for inclusion in the systematic review is thoroughly assessed. The inclusion/exclusion criteria applied strictly during the screening process, with reasons for exclusion documented for transparency and reproducibility·

\subsection{Data Extraction:}
Following the retrieval of results from the database search, the identified records were imported into a reference management tool, EndNote, to organize and consolidate the search outcomes. During this process, duplicate entries were systematically identified and removed. The subsequent screening of studies was conducted independently by two reviewers, Alzahra Altalib (AA) and Alessandro Perelli (AP), who applied the predefined eligibility criteria to determine the suitability of studies for inclusion in the systematic review. Any disagreements between the reviewers were resolved through discussion to achieve consensus. Data extraction was performed from the final set of selected articles using a standardized approach. Key details were collected, including publication information (e.g., year, authors, and country of origin) and specific parameters related to diffusion models, such as noise injection methods, dataset types, and model architectures etc.

\subsection{Quality Assessment:}
To assess the methodological quality of the included studies, two researchers, AA and PA, employed the Quality Assessment of Diagnostic Accuracy Studies-2 (QUADAS-2) tool, as outlined by (\citet{reitsma2012accuracy}). This tool was specifically used to evaluate the risk of bias and ensure methodological rigor. By providing a structured framework, QUADAS-2 facilitated a systematic assessment of potential biases, the overall quality, and the robustness of each study included in the systematic review.

\subsection{Data Synthesis:}
The synthesis data was examined and presented to identify general patterns, advantages, limitations, and deficiencies in the research pertaining to the use of diffusion models for sCT generation. The information extraction was based on pre-defined criteria that have been presented later in Table \ref{summary}.

\section{Results}\label{sec:results}
\begin{figure}[!h]
    \centering
    \includegraphics[width=\linewidth]{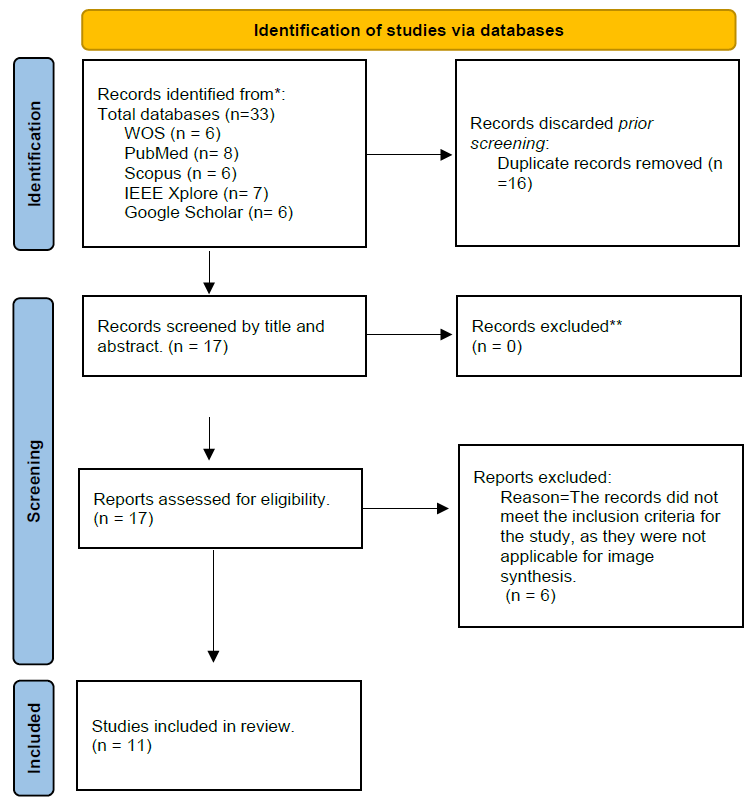}
    \caption{Standard workflow of Conditional Diffusion models for CBCT to CT image synthesis.}
    \label{fig:PRISMA_flow}
\end{figure}

The database search yielded a total of 33 records, distributed as follows: 6 from Web of Science (WoS), 8 from PubMed, 6 from Scopus, 7 from IEEE Xplore, and 6 from Google Scholar. After the removal of duplicate entries, 17 unique records remained. These records were then assessed based on the predetermined inclusion and exclusion criteria. As a result, 6 records were excluded because they did not meet the study's inclusion criteria, specifically as they were not applicable to image synthesis. Consequently, 11 records were included in this systematic review. The screening process is comprehensively outlined in PRISMA Fig. \ref{fig:PRISMA_flow}.

The articles included and reviewed as part of this study collectively explored advanced methods for the sCT generation using CBCT and CT images. In all of these reviewed studies, emphasis has been kept on the use of the diffusion model to achieve images with improved quality, artifacts reduction, and be suitable for the clinical settings. Numerous approaches have been presented including frequency-guided diffusion models (FGDM) (\citet{li2023fgdm}), stacked coarse-to-fine architectures (\citet{sun2024pseudo}), and patient-specific fine-tuning (\citet{chen2024cbct}, \citet{peng2024conditional}). These works propose to address the inherent limitations associated with CBCT images such as noise, artifacts and inaccuracies associated with Hounsfield Unit (HU) values. The studies incorporated a diverse range of datasets that included paired and unpaired CBCT-CT image slices, dual-energy CT (DECT) scans etc. These scans have been developed for the radiotherapy and proton therapy context (\citet{viar2024dual}). The study further incorporated numerous loss functions including frequency-domain regularization (\citet{li2023fgdm}), edge-aware constraints (\citet{zhang2024texture}), and hybrid condition losses (\citet{peng2024conditional}, \citet{sun2024pseudo}). These loss functions help in preserving anatomical details although  achieving structural and dosimetric accuracy. The proposed neural network architectures, including Swin-UNETs (\citet{viar2024dual}), dual-branch attention modules (\citet{zhang2024texture}), and texture-preserving frameworks (\citet{zhang2024texture}) have shown to achieve improved performance in terms of PSNR, MAE, and SSIM, and have been found to outperforming traditional GAN and VAE-based methods (\citet{li2023fgdm}). Despite these advancements, challenges such as computational inefficiency, dependency on paired datasets, and limited robustness to frequency variations persist (\citet{li2023fgdm}, \citet{sun2024pseudo}, \citet{zhang2024texture}). Overall, the findings highlight the transformative potential of diffusion-based models for improving CBCT-to-CT synthesis. The studies have found improvements in adaptive radiotherapy, proton therapy planning, and broader clinical applications (\citet{chen2024cbct}, \citet{li2023fgdm}, \citet{viar2024dual}. A detailed synthesis of the included articles has been presented in Table \ref{summary}

\section*{Overview of Diffusion Model Families for CT Image Synthesis}

\renewcommand{\arraystretch}{1.6}

\begin{sidewaystable}
\centering
\caption{Synthesis of the Included Articles on sCT generation using Diffusion Models.}\label{summary}
\resizebox{\textwidth}{!}{\small
\begin{tabular}
{p{65pt}p{65pt}p{65pt}p{65pt}p{75pt}p{75pt}p{55pt}p{55pt}p{55pt}p{50pt}p{55pt}p{55pt}}
\hline

\textbf{Author / Year} & \textbf{Study Type} & \textbf{Population} & \textbf{Pre-processing
}  & \textbf{Conditional Loss Function (Reverse Process) } & \textbf{Type of Neural Network (Reverse Process)} & \textbf{Training Strategies} & \textbf{Data Type)} & \textbf{Outcomes (MAE, PSNR, etc.)} & \textbf{Noise Injection (Type and Levels)}& \textbf {Limitation} & \textbf{Findings} 
\\
\hline
(Li et al., 2023 \citet{li2023fgdm}) &Frequency-Guided Diffusion Model (FGDM) & CBCT-CT datasets across institutions (various samples) & Frequency domain filtering (high-pass and low-pass) & Frequency-guided regularization & Freq.-guided diffusion model & Zero-shot domain adaptation & Paired and unpaired CBCT-C T data & FID improved, PSNR: 30+ dB & Controlled Gaussian noise & Limited robustness to freq. domain changes & Preserves structure in translation \\
\hline
(Chen, Qiu, Peng, et al., 2024) \citet{chen2024cbct} & Patient-specific model & Lung cancer (33 pts) & Normalization, fine-tuning & Fine-tuned with patient data & General lung diffusion model & Paired patient-specific training & Paired 2D slices & MAE: 15.96 HU, PSNR: 33.57 dB & Gaussian noise augmentation & Time-intensive tuning & Improved sCT quality, artifact correction \\
\hline
(Chen, Qiu, Wang, et al., 2024) \citet{chen2024patient} & Patient-Specific Diffusion Model & Lung Cancer Patients (33 patients) & Anatomical fine-tuning & Custom fine-tuning loss & Patient-Specific DDPM &Paired fine-tuning per patient & Paired 2D slices & MAE: 15 HU, PSNR: 33 dB & Gaussian noise  & High computation per patient & Effective sCT improvement \\
\hline
(Fu et al., 2024) \citet{fu2024energy} & Diffusion model study & Chest Tumor Dataset (100+ samples) & Normalization & Energy-guided loss & UNet & Markov Chain Sampling & Unpaired 3D slices & MAE: 26.87 HU, PSNR: 19.83 dB & Gaussian noise & Mode collapse in GAN comparison & Superior to GAN-based methods \\
\hline
(Li et al., 2024) \citet{li2024zero} & FGDM & CBCT-CT datasets across institutions (various) & Frequency domain analysis & Frequency-guided regularization & Frequency-guided diffusion mode & Zero-shot domain adaptation & Paired and unpaired CBCT-CT data & FID improved, PSNR: 30+ dB & Frequency-guided noise handling & Limited robustness to frequency domain changes & Preserves structural details during domain translation \\
\hline
(Peng, Gao, et al., 2024)\citet{peng2024unsupervised} & Unsupervised Bayesian Framework & H\&N, Lung, Pancreas (75 pts) & Score-based patient-specific priors & Total variation regularization & Patient-specific diffusion model & Unsupervised patient-specific training & Unpaired 3D slices & MAE: 50 HU, PSNR: 31 dB & Score-based noise scheduling & Slice alignment sensitivity & Effective artifact reduction \\
\hline

\end{tabular}
}
\end{sidewaystable}

\begin{sidewaystable}
\centering
\caption{Synthesis of the Included Articles on sCT generation using Diffusion Models.}\label{summary}
\resizebox{\textwidth}{!}{\small
\begin{tabular}
{p{65pt}p{65pt}p{65pt}p{65pt}p{75pt}p{75pt}p{55pt}p{55pt}p{55pt}p{50pt}p{55pt}p{55pt}}
\hline

\textbf{Author / Year} & \textbf{Study Type} & \textbf{Population} & \textbf{Pre-processing
}  & \textbf{Conditional Loss Function (Reverse Process) } & \textbf{Type of Neural Network (Reverse Process)} & \textbf{Training Strategies} & \textbf{Data Type)} & \textbf{Outcomes (MAE, PSNR, etc.)} & \textbf{Noise Injection (Type and Levels)}& \textbf {Limitation} & \textbf{Findings} 
\\
\hline

(Peng, Qiu, et al., 2024)\citet{peng2024conditional} & Conditional model & Brain, H\&N (50 pts) & Cropping, normalization & L2 loss & Time-embedded UNet & Paired image data & Paired 2D slices & MAE: 25.99 HU, PSNR: 30.49 dB & Gaussian noise (time steps) & Requires large paired data & Improved CBCT quality for ART \\
\hline
(Sun et al., 2024)\citet{sun2024pseudo} & Stacked coarse-to-fine & Pelvic cancer (250 pts) & Multi-stage denoising & Edge-preserving loss & DDPM with U-ConvNeXt & Hierarch training & Paired CBCT-CT slices & PSNR: 34.02 dB, SSIM: 87.14\% & Hierarchical denoising & Paired data dependency & Enhanced ART dosimetric accuracy \\
\hline
(Viar-Hernandez et al., 2024)\cite{viar2024dual} & Dual Energy Synthesis & H\&N (54 pts) & CBCT-DECT normalization & Gradient matching loss & Multi-decoder Swin-UNET & Multi-decoder learning & Paired CBCT-DECT slices & MAE: 39.58 HU, PSNR improved & Controlled Gaussian noise & Dual-energy data complexity & Improved tissue characterization \\
\hline
(Yin et al., 2024)\cite{yin2024hc3} & Latent diffusion model & Prostate cancer (30 pts) & FFT-based high-freq extraction & Hybrid condition loss & Unified feature encoder & Hybrid high-freq embedding & Paired 3D slices & Gamma Passing Rate: 93.8\% & Gaussian noise with FFT & Computation inefficiency & Enhanced anatomical preservation \\
\hline
(Zhang et al., 2024)\cite{zhang2024texture} & Texture-preserving model & Multicenter CBCT-CT (100+ pts) & FFT, wavelet transforms & Boundary-aware loss & Dual-branch attention model & High-freq optimization & Unpaired 3D volumes & MAE: 18.48 HU, PSNR: 33.07 dB & Adaptive high-freq noise handling & High compute demand & Superior texture preservation \\
\hline

\end{tabular}
}
\end{sidewaystable}

\subsection{Quantitative Results}
\begin{figure}[!h]
    \centering
    \includegraphics[width=.85\linewidth]{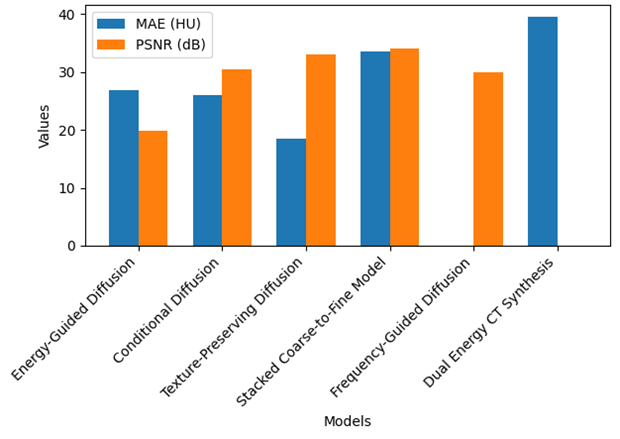}
    \caption{MAE and PSNR achievement for the Reviewed Models}
    \label{fig:PSNR}
\end{figure}

The quantitative analysis of the articles reviewed has been carried out as part of this section where three measures are identified: performance metrics, dataset characteristics, and improvements achieved using the diffusion models for sCT generation. 
For the performance metrics assessment, the reported performance metrics including MAE, and PSNR across various diffusion models have been analyzed. It can analyzed in Fig. \ref{fig:PSNR} that MAE and PSNR across diffusion-based models are significant. For example, the texture-preserving diffusion model has achieved an MAE of 18.48 HU and a PSNR of 33.07 dB depicting high-quality image achievement. Similarly, the conditional diffusion model has shown a balanced performance in terms of both the MAE and PSNR. The outcomes indicate the diffusion models have been found to achieve improved performance by reducing the artifacts associated with CBCT data and by improving the structural fidelity.

\begin{figure}[!h]
    \centering
    \includegraphics[width=.85\linewidth]{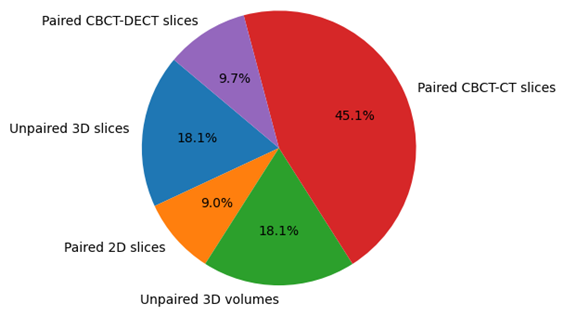}
    \caption{Data sampling and distribution across models.}
    \label{fig:Data Sampling}
\end{figure}

The studies have shown that the dataset used by the studies varies in size and type as depicted in Fig. \ref{fig:Data Sampling}. This helps with analysing these models in the context of several clinical settings. For instance, the stacked coarse-to-fine model has made use of paired CBCT-CT dataset for 250 pelvic cancer patients. This explains its applicability to large and domain-specific datasets. Frequency Guided diffusion model has shown its versatility by showcasing an ability to handle both paired and unpaired datasets. Such scalability helps models to achieve high performance even when the datasets are small For instance, the 50-patient brain and H\&N dataset has been used in the conditional diffusion.

\begin{figure}[!h]
    \centering
    \includegraphics[width=.85\linewidth]{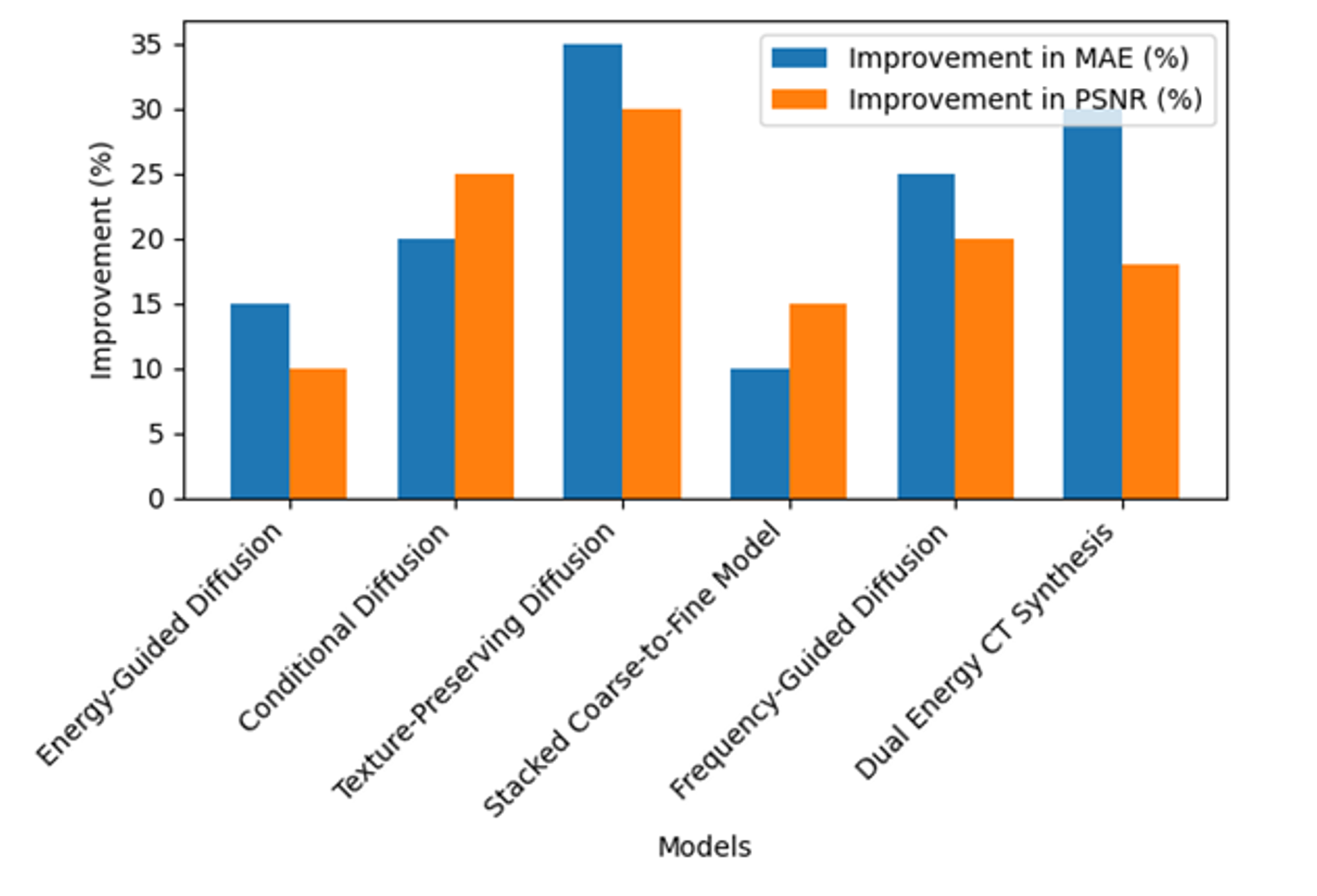}
    \caption{Percentage improvement in the model performance compared to traditional methods.}
    \label{fig:Percentage Improvement}
\end{figure}

Finally, a comparison in terms of performance improvement compared to traditional models recorded by the studies has been depicted in Fig. \ref{fig:Percentage Improvement}. Diffusion models have been found to improve the MAE and PSNT performance compared to GANs and VAEs. The Texture-Preserving Diffusion model has been found to achieve an improvement of 35\% in MAE and a 30\% gain in PSNR. Similarly, Frequency-Guided Diffusion has achieved a 25\% improvement in MAE and a 20\% gain in PSNR.

\section{Discussion}
The findings from the review highlight the potential of conditional diffusion models in sCT generation. The diffusion models have been found to address the challenges associated with low-quality CBCT images in the form of artifacts, noise, and structural inaccuracies. The use of diffusion models is relatively unexplored, however the reviewed models include FGDM, Texture-Preserving Diffusion Models, and Stacked Coarse-to-Fine Models as being the most dominant and promising methods. These methods reside on the use of advanced loss functions such as frequency-domain regularization and edge-aware constraints. This helps them to retain the fine anatomical details and to improve the dosimetric accuracy. The use of datasets across the studies further highlights the adaptability of the models across several clinical settings. The dataset approach ranges from single-institution paired datasets and multi-center unpaired datasets. The FGDM and Texture-Preserving Diffusion have further illustrated the MAE and PSNR performance and were found to improve sCT image quality by reducing the artifacts. 
Compared to the traditional deep learning models including GANs and VAEs, the diffusion models are outperforming. This is valid in terms of structural fidelity and quantitative metrics. For example, the Texture-Preserving Diffusion model has shown an MAE improvement of 35\% and a PSNR improvement of 30\%. This is potentially due to the iterative refinement approach adopted by the models that render high-quality images by integrating domain-specific priors and noise reduction. However some of the challenges remain to be addressed including the computational requirement and high reliance on the paired datasets. This may somehow limit them to becoming ubiquitous in clinical settings. 
In clinical settings, the use of diffusion models for sCT generation can have significant implications in terms of adaptive radiotherapy and proton therapy planning. The models work by reducing the artifacts and improving HU accuracy thereby leading to improved dose computations and precise treatment planning. This is especially true for the anatomically challenging regions. Models like FGDM can work on the unpaired datasets to facilitate scalability as well.

\subsection{Answer to Research Questions}
The three research questions identified as part of this work in the introduction have been answered as follows: 
\begin{enumerate}
    \item The conditional diffusion models for sCT generation are limited yet have adopted several approaches. These include Frequency-Guided Diffusion Models, Texture-Preserving Diffusion Models, and Stacked Coarse-to-Fine Models. The models are based on advanced loss functions including frequency-domain regularization, edge-ware constrained and hybrid loss conditional losses. This helps with ensuring that high anatomical accuracy is achieved and artifacts are reduced. Additionally, some adoptive techniques like dual-mode feature fusion and hierarchical learning are also found in the articles reviewed. 
    \item The diffusion models have typically been found to outperform the traditional deep learning models. These specifically include GANs and VAEs in terms of accuracy. The quantitative assessment has shown that the models consistently achieve low MAE and high PSNR. The best-performing model Texture-Preserving Diffusion has shown a 35\% improvement in MAE and 30\% improvement in PSNR performance compared to GANs. Such improvements are due to the adoption of an iterative approach using domain-specific priors. This allows diffusion models to handle the noise and artifacts effectively.  
    \item The clinical implications of diffusion models for sCT generation are expected to be numerous. By enabling a reduction of artifacts and showing a tendency to increase HU accuracy, diffusion models can allow precise dose calculations. This may lead to improved treatment planning in adaptive radiotherapy and proton therapy. The ability of the models to work with unpaired datasets (FGDM for instance) helps with improved scalability and thus applicability in many clinical settings. Overall, diffusion models have shown the potential to improve patient outcomes through safe and effective radiation treatments. However, further validations are needed to analyze the computational inefficiencies associated with these models. 
\end{enumerate}

\subsection{Risk of bias assessment}
\begin{figure}[!h]
    \centering
    \includegraphics[width=\linewidth]{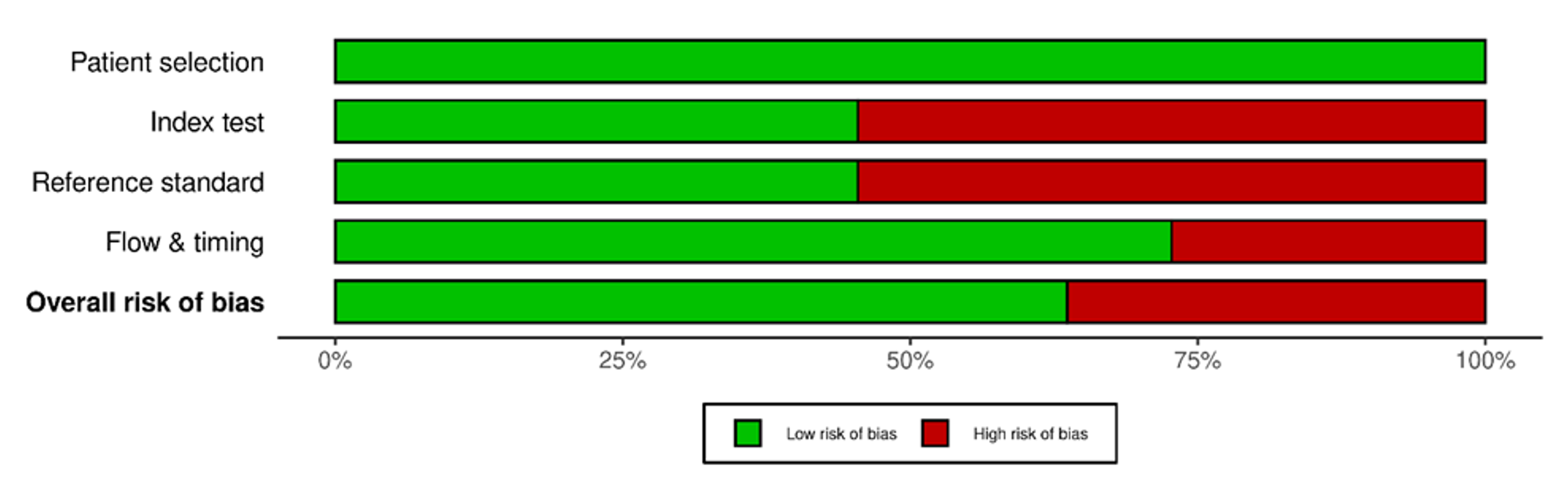}
    \caption{The risk of bias assessment results for all the included studies were conducted using QUADAS 2.}
    \label{fig:The risk of bias assessment}
\end{figure}

The QUADAS-2 assessment revealed variations in the methodological quality of studies and a risk of bias. Although  all studies had "Low" risk in Patient Selection, only \citet{fu2024energy}, \citet{zhang2024texture}, and \citet{viar2024dual} achieved "High" overall ratings, indicating strong validation and reliable reference standards. In contrast, studies like \citet{sun2024pseudo} and \citet{peng2024unsupervised} were rated "Low" overall due to inadequate validation and unclear flow. Overall, these findings have been depicted in Fig. \ref{fig:The risk of bias assessment}.

\subsection{Registration and Reporting}
The findings of this systematic review were reported in adherence to the Preferred Reporting Items for Systematic Reviews  (PRISMA) guidelines. The review was conducted in accordance with a pre-established protocol registered with the International Prospective Register of Systematic Reviews (PROSPERO), bearing the registration number CRD42024619240.

\section{Conclusion}       
In conclusion, the advancement in the diffusion models including conditional and denoising diffusion approaches has been found to exhibit high performance for sCT generation from CBCT images. The model helps bridge the gaps with the traditional models in terms of noise handling and achieveing high structural fidelity. The diffusion models have been found to outperform the traditional models including GANs and VAEs by exhibiting higher MAE, PSNR, and SSIM performance. These models reside on an iterative refinement approach with domain-specific priors. This enables them to extract accurate image synthesis as well as dose calculations and treatment planning. This is especially true in adaptive radiotherapy and proton therapy. Despite these benefits, some of the validations need to be carried out including the analysis of computational inefficiencies and experimentation of real-world clinical data (especially unpaired). Future research shall focus on the optimization of these models for their clinical scalability and to ensure their robust performance in the inter-subject domain. In summary, diffusion models are found to hold promising outcomes in radiotherapy outcomes and have the potential to improve patient care through precise and reliable imaging.

\section{Acknowledgement:}
This study did not involve the use of human subjects or animals in its research. The authors declare no conflicts of interest, whether financial or personal, that could influence or be relevant to the work presented in this paper. A. Perelli acknowledges the support provided by the Royal Academy of Engineering through the RAEng/Leverhulme Trust Research Fellowships program (award number LTRF-2324-20-160).

\bibliographystyle{elsarticle-num-names} 
\bibliography{biblio}

\end{document}